# CompARE: A Computational framework for Airborne Respiratory disease Evaluation integrating flow physics and human behavior

Fong Yew Leong[1+], Jaeyoung Kwak[2+], Zhengwei Ge[1], Chin Chun Ooi[1,3]*, Siew-Wai Fong[4], Matthew Zirui Tay[4], Hua Qian[5], Chang Wei Kang[1], Wentong Cai[2], Hongying Li[5*]

[1] Institute of High Performance Computing, Agency for Science, Technology and Research, 1 Fusionopolis Way, Singapore 138632, Singapore

[2] College of Computing and Data Science, Nanyang Technological University, 50 Nanyang Avenue, Singapore 639798, Singapore

[3] Centre for Frontier AI Research, Agency for Science, Technology and Research, 1 Fusionopolis Way, Singapore 138632, Singapore

[4] A*STAR Infectious Diseases Labs, Agency for Science, Technology and Research, 8A Biomedical Grove, Singapore 138648, Singapore

[5] School of Energy and Environment, Southeast University, Nanjing 210096, China

[6] School of Mechanical and Aerospace Engineering, Nanyang Technological University, 50 Nanyang Avenue, Singapore 639798, Singapore

[+] Authors contributed equally; *co-corresponding author

Corresponding author e-mail address: ooicc@cfar.a-star.edu.sg

**Abstract**

The risk of indoor airborne transmission among co-located individuals is generally non-uniform, which remains a critical challenge for public health modelling. Thus, we present CompARE, an integrated risk assessment framework for indoor airborne disease transmission that reveals a striking bimodal distribution of infection risk driven by airflow dynamics and human behavior. Combining computational fluid dynamics (CFD), machine learning (ML), and agent-based modeling (ABM), our model captures the complex interplay between aerosol transport, human mobility, and environmental context. Based on a prototypical childcare center, our approach quantifies how incorporation of ABM can unveil significantly different infection risk profiles across agents, with more than two-fold change in risk of infection between the individuals with the lowest and highest risks in more than 90% of cases, despite all individuals being in the same overall environment. We found that infection risk distributions can exhibit not only a striking bimodal pattern in certain activities but also exponential decay and fat-tailed behavior in others. Specifically, we identify low-risk modes arising from source containment, as well as high-risk tails from prolonged close contact. Our approach enables near-real-time scenario analysis and provides policy-relevant quantitative insights into how ventilation design, spatial layout, and social distancing policies can mitigate transmission risk. These findings challenge simple distance-based heuristics and support the design of targeted, evidence-based interventions in high-occupancy indoor settings.

1. Introduction

Respiratory diseases like COVID-19 spread primarily through the emission of infectious respiratory particles from activities such as coughing, sneezing, or talking, that can infect individual across a range of distances and timescales, depending on environmental conditions.[1,2] Respiratory transmission can facilitate rapid disease spread, particularly in areas where large numbers of people congregate, including schools, workplaces, and public transportation.[2,3] Environmental factors such as human mobility,[4,5] air flow in buildings,[6–8] and public health measures[9] can either mitigate or contribute towards spread of respiratory diseases. Understanding these factors are crucial for controlling the spread of respiratory diseases and reducing their impact on public health.

Computational fluid dynamics (CFD) is a useful method for modeling air flows, which play a crucial role in respiratory disease transmission, especially in indoor environments. Since the

COVID-19 outbreak, CFD has been employed extensively for infection risk assessment under various environmental conditions, indoors and outdoors.[10–14] In view of the COVID-19 pandemic, various organizations including World Health Organization (WHO) and the Center for Disease Control and Prevention (CDC) have provided guidance and recommendations for improving ventilation in various settings. The American Society of Heating, Refrigerating and Air-Conditioning Engineers (ASHRAE) has also developed guidelines and standards for ventilation and indoor air quality in buildings.[15]

Environmental factors aside, human mobility, including both local and long-distance movement, plays a significant role in the spread of respiratory infectious diseases such as COVID-19. Various models can be used to study human mobility including agent-based model (ABM),[16] gravity model,[17,18] network model[19] and information-theoretic model.[20] In particular, ABM has been successfully deployed to inform policy planning during the pandemic. Typical CFD models do not consider human mobility within simulations, and models such as ABM need to be incorporated to account for human mobility. CFD models the air flow patterns in an idealized recreation of the physical built environment, whereas ABM accounts for the diverse behavioral choices and interactions between individuals, making the insights from the two methodologies very complementary.[21] Hence, coupling of both CFD and ABM models to incorporate human mobility has been studied as a way to provide more realistic and helpful predictions and guidance in the context of evacuation planning,[16] emergency preparedness,[15,20,22–24] and the design of safer built environments.[15,22–25] Epstein et al.[16] initially proposed coupled CFD-ABM model for urban evacuation planning. The study considered how airflow dynamics, smoke propagation, building layout as well as individual behaviors and interactions affect evacuation. Similar methodology was adopted by others to investigate crowd dispersal due to chemical or toxic gas exposure,[15,22,23] and cells on microcarriers in a stirred-tank bioreactors.[24,25]

Interestingly, little work has been done in current literature on applying a CFD-ABM framework to the study of airborne infectious diseases such as COVID-19. Schinko et al.[26] used a coupled CFD-ABM model for an airborne transmission study in a supermarket scenario, where the ABM emulated agents following decisions according to a goal sequence based on behavior of past in-store shoppers, namely, Enter, Shop, Queue, Pay, and Exit, in that order. These agents are mobile and perform aerosol-emitting actions in stochastic fashion, such as breathing, coughing, and talking.[26] However, the study remained limited by the resources required for extensive simulations of such large-scale scenarios or environments with intricate details, even with the adoption of GPU-accelerated Lattice Boltzmann Method.

More generally, while CFD can provide detailed insights into the underlying flow physics, and ABM can capture individual-level interactions for more comprehensive and complementary analysis and insights, the requisite computational cost can hinder wide-spread integration. CFD is renowned for its high computational cost especially when repeated calculations are required for parametric studies involving various flow and dispersion scenarios.[27] These computational demands become prohibitive when CFD is coupled with ABM, as may be required to capture changes such as displacements in emission source due to human activity. In recent years, advancements in machine learning (ML) have demonstrated considerable potential in mitigating these challenges, successfully accelerating CFD simulations across multiple applications.[28–33] These innovations not only improve efficiency but also enable the exploration of a broader range of scenarios involved with parametric study within a shorter time frame. Hence, the ML-enabled integration of CFD and ABM represents another transformative opportunity to enhance risk analysis in respiratory disease transmission.

Inspired by the potential for ML to significantly enhance computational efficiency and ease of implementation, we propose a novel integrated modeling framework that leverages these three modeling approaches to assess the transmission risks of respiratory diseases. The framework operates in three stages. First, CFD is used to predict airflow and aerosol distribution in an indoor environment. Next, a surrogate ML model is trained on the CFD data to rapidly predict spatial aerosol concentrations. These ML-derived aerosol distributions are then integrated into ABM simulations, which model human mobility and interactions. Finally, within this compositional framework, the probability of infection for all individuals is quantitatively assessed in a simple and flexible manner.

To demonstrate the practical utility of this integrated framework, we conducted a case study simulating a prototypical childcare center in Singapore. Our analysis highlights the synergistic value of combining all three methodologies, with particular emphasis on how human mobility patterns critically influence airborne transmission risks. To the best of the authors' knowledge, this is the first integrated methodology that combines CFD, data-driven modeling, and human mobility analysis. Critically, this integration also represents a fundamental advance over existing approaches, enabling a more comprehensive, multi-factorial quantification of the second order statistics of the probability of infection individuals face due to the complex interplay between environmental airflow patterns, human activity and disease biology. The proposed framework offers a powerful tool for acquiring insights to enhance practical public health interventions and evidence-based indoor safety management.

## 2. Methods

### 2.1. Modified Wells-Riley approach to multi-factorial environmental and human activity-aware assessment of probability of infection for airborne infectious disease

Respiratory infectious diseases, exemplified by the SARS-CoV-2 pandemic, can be spread from an infected (even asymptomatic) host to others through inhalation of emitted droplets or nuclei across time and space.[34] Airborne infection risk assessment is carried out in two main steps:[35]

1. *Transport* – The infection dosage is estimated based on exhalation, airborne dispersion and inhalation of infectious respiratory particles.
2. *Risk* – The infection risk is estimated based on infection dosage using a disease-specific dose-response relationship.

Building on previous work by Mittal et al.,[34] the *Transport* step can be decomposed into 3 key processes, where we model (i) the emission from the infected person, (ii) the transport and dispersion of infectious quanta from the infected individual(s) to the susceptible individual(s), and (iii) the total biological exposure (e.g. through inhalation) the susceptible individual has accumulated within the entire duration of the scenario. We determine the infectious quanta emission rate from (i) and extent of dispersion from (ii) and incorporate them in quantitative models such as the well-known Wells-Riley model to estimate the risk of infection.

The Wells-Riley model[36,37] quantifies risk using the concept of an infectious quantum, defined as the inhaled dose of airborne viruses required to infect 63.2% of susceptible persons in an enclosed space based on Poisson statistics. The probability of infection $P_j$ for a susceptible person $j$ can be estimated as,

$$P_j(\tau) = 1 - e^{-Q_j(\tau)}, \tag{1}$$

where $Q_j(\tau) = ER_q \cdot \int_0^\tau k_t(t)\, dt$ is the cumulative dose received over time $\tau$ and $ER_q$ is the infectious quanta emission rate from the infected person (in units of quanta per unit time). $k_t(t)$ is a transport scaling factor that accounts for the spatial dispersion of infectious quanta emitted from the infected source $i$ located at $\vec{x}_i(t)$, and pulmonary intake from the susceptible individual $j$ located at position $\vec{x}_j(t)$ at time $t$. Accurate risk assessment therefore hinges on obtaining a meaningful estimation of the related emission quantities, $ER_q$, and the transport scaling factor, $k_t(t)$.

Under the commonly employed well-mixed assumption in indoor environments, $k_t$ is approximated by key ventilation parameters such as the air change per hour being provided to the environment. However, over the past few years, there have been numerous efforts towards better quantification of this transport parameter $k_t$ using high-fidelity numerical simulations such as CFD and/or other theoretical models.[35,38,39] While physical models can account for environmental factors, $k_t$ is actually a factor of the interaction dynamics between the infected and susceptible individual, a complexity which is best captured in combination with other simulation methodologies such as agent-based models.

In the following subsections, we detail a new risk assessment framework which utilizes a coupled CFD-ML-ABM methodology (illustrated in **Figure 1**). Under the ABM framework, time is discretized in sequential clearance intervals termed as "breath events" (**Appendix D**), so that

$$Q_j(\tau) = ER_q \cdot \beta \cdot \sum_{i=1}^{n_\beta} k_i, \qquad (2)$$

where $\beta$ is a clearance period (in units of time), estimated here as 1/15 min per breath for normal breathing. $k_i$ is the transport scaling factor within a clearance interval $i$ and summed to total number of intervals $n_\beta = \tau/\beta$. Enabled by ML-based acceleration within an extended Well-Riley model, we demonstrate its applicability towards understanding airborne infectious disease in the context of a prototypical childcare center (described in greater detail in **Section 3.1**).

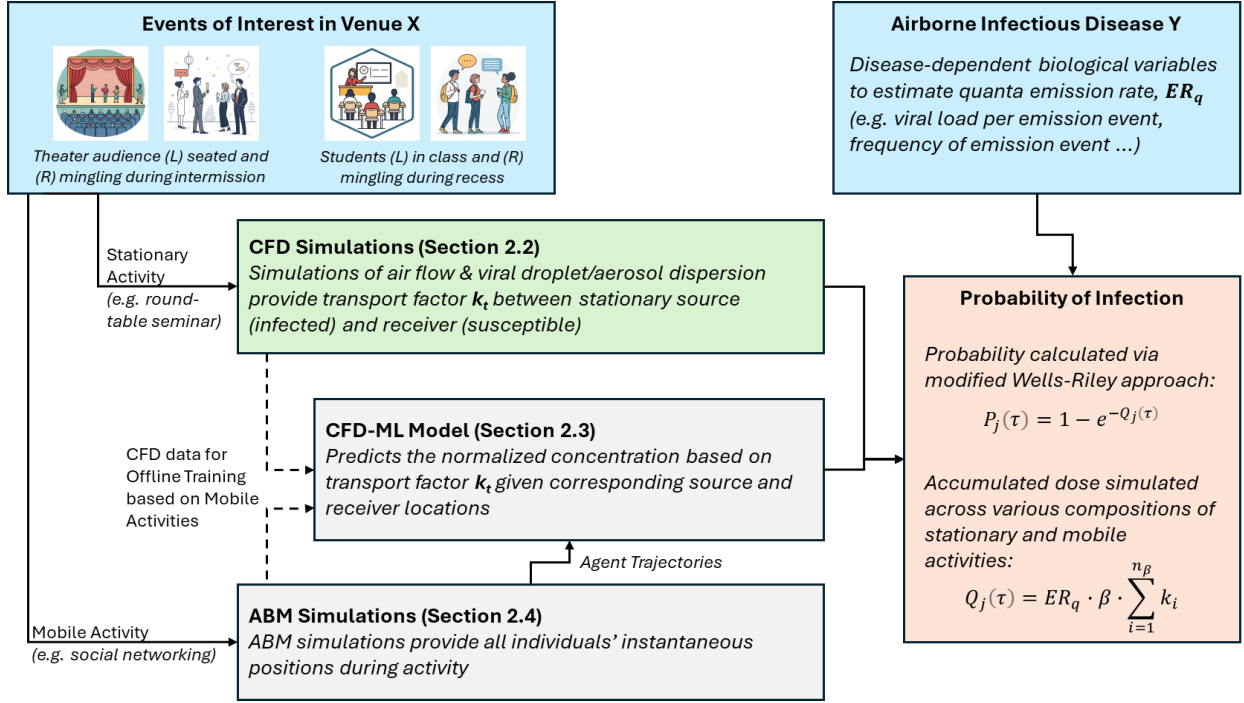

**Figure 1**. Schematic of the workflow in the proposed modified Wells-Riley approach to a comprehensive, multi-factorial framework for assessment of probability of infection for airborne infectious disease that integrates environmental, human activity and disease biology.

## 2.2. CFD transport model

Consider infectious respiratory particles emitted continually by an infected source and carried by the air circulating in the enclosed building. Over time, the droplet concentration in the air develops and approaches steady state, presenting calculable infection risk via inhalation to susceptible persons in the venue. Here we employ CFD to solve the governing equations, including mass, heat and aerosol transport equations, using *k-ε* turbulence model with scalable wall functions (**Appendix A**). The computational domain encompasses the lobby area, classroom space, and the adjacent washroom within a prototypical childcare centre (as described in **Section 3.1** and **Appendix B**).

## 2.3. Machine Learning (ML)

As numerical simulation is computationally expensive, and key steps such as meshing to place a spatially-temporally-varying source box in the computational domain (to simulate emission from the infected individual) can be time-consuming, we develop surrogate models to predict the spatial concentration of a passive scalar dispersed within an indoor environment, in this case the childcare center. The training data utilized in the surrogate models were derived from offline-derived high-fidelity numerical simulations of the passive scalar equation within the indoor environment acquired in combination with the flow simulation.

We assume instant dispersion and clearance under steady state conditions (**Appendix D**). Then, the model's output is a single value corresponding to the normalized passive scalar concentration at a specific location. Under the simulated release of a normalized viral quantity from the infectious individual (source), the corresponding concentration at the susceptible individual's location (receiver's location) also corresponds to the instantaneous $k_t(t)$ described above. Hence, the model prediction corresponding to ABM-derived source and receiver locations allow us to calculate the cumulative inhaled quanta through an entire scenario.

We tested two distinct surrogate modelling techniques, namely XGBoost and Multi-layer Perceptrons (MLPs) (**Appendix E**). Both models were trained using ($x$,$y$) positional inputs of the infectious agent (source) and the susceptible agent (receiver), and the horizontal Euclidean distance between them. While deep learning is popular in machine learning applications, simpler surrogate models such as XGBoost (a regression tree model) still perform well. This is likely because the dataset size (100 scenarios) is relatively small, which may not fully leverage the benefits of deep learning approaches. In general, the surrogate models are extremely fast and, once trained, can provide predictions in near real-time (within a second). Both surrogate models performed well in our testing. The XGBoost model was selected for its ease of integration, as it is lightweight and can be easily coupled with the ABM models (**Appendix F**).

## 2.4. Agent-based Model (ABM)

To model human mobility, we develop an ABM based on a hierarchical behaviour modelling framework, which consists of the strategic layer, the tactical layer, and the operational layer.[40,41]

The strategic layer determines destinations for individuals, while the tactical layer is associated with how the individuals approach their destinations. We assigned destinations for individuals based on an origin-destination (OD) matrix, which defines the probability that individuals would move from one area to another area. When the break begins, all individuals can use the washroom, join a subgroup, or stay at their table seats. Also, the individuals that have used the washroom will either return to the table or join a subgroup. Over time (end of the break), individuals will gradually leave the subgroups and return to the table. As the venue size is small, all individuals have complete knowledge of the walking environment, so we applied Djikstra's algorithm[45] to mimic individuals' behaviour of finding the shortest route to the destinations (tactical layer). Lastly, the operational layer describes how the individuals make actual stepwise movements such as walking from point to point (**Appendix F**).

## 2.5. Risk assessment

Given the passive scalar ML model and ABM model outputs of potential agent trajectories, we can compute the probability of infection for susceptible individuals over each single simulation scenario. In addition to transport scaling factor $k_t$, the quanta emission rate $ER_q$ is required to estimate $P_j(\tau)$, the infection probability of a susceptible agent $j$ over time $\tau$ (**Section 2.1**). Assuming a walking, talking infected source, the mean quanta emission rate for COVID-19 is estimated as $ER_q \approx 0.232$ (q/min), based on literature sources (**Appendix G**). Using our CFD-ML-ABM model, we estimate $k_i$ for a clearance interval $i$ between pairs of infectious and susceptible individual, which allow us to estimate each susceptible individual's probability of infection based on the cumulative dose $Q_j(\tau)$ within the period of interest $\tau$.

## 3. Results

Childcare centers are a perennial concern for the spread of infectious disease, especially because of their dense concentration of vulnerable persons (the very young). During March 2020, there was also a significant COVID-19 outbreak in Singapore at a childcare center with many members of the staff being infected. Given the complexity and variety of activities occurring in such locations, this study seeks to demonstrate the capability of this framework to provide insights into the potential risk of airborne infection across multiple activities.

Here, we present results for the CFD-ML-ABM model framework in that order, including dispersion of aerosols and movements of individuals. This is followed by a quantitative risk assessment, where we track the total dose incurred by an individual throughout the exposure period to estimate the probability of infection.

## 3.1. Case study on childcare center

The computational model encompasses the lobby area, classroom space, and the adjacent toilet, as shown in **Figure 2**. The classroom is equipped with six ceiling-mounted air-conditioning cassettes shown as dark blue colour, with an open doorway connecting it to the toilet. Two toilet exhaust fans are in continuous operation to expel air from the restroom. During normal operations, the main door to the lobby area is closed, hence the primary ventilation into the office area is via the air-conditioning units.

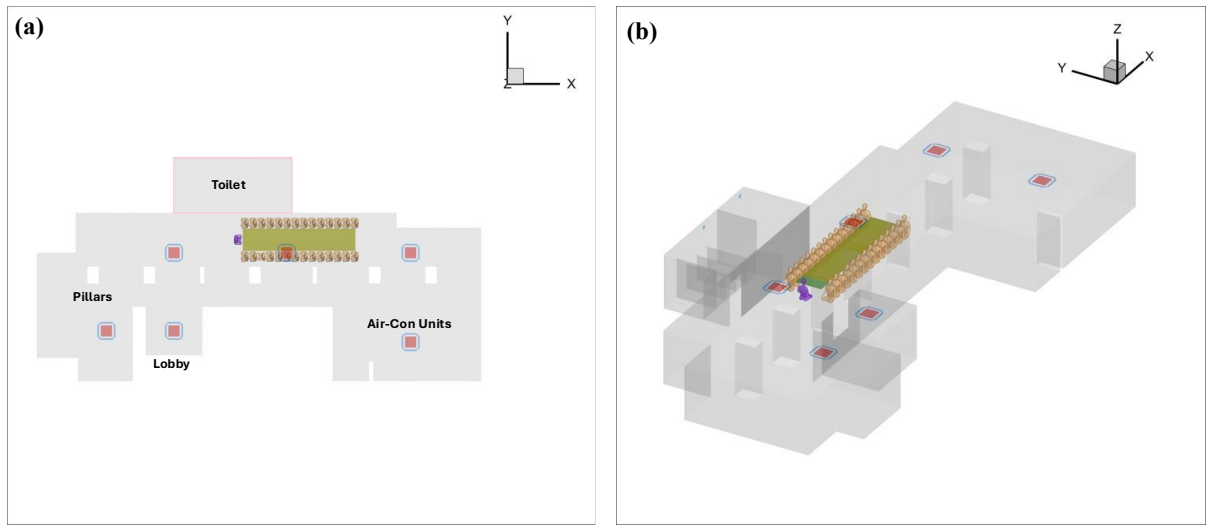

**Figure 2.** (a) Top down and (b) iso-geometric views of the childcare center. The central area is equipped with air-conditioning units on the ceiling (blue and pink squares) and a long table flanked by chairs where staff sit. Other accessible locations include the toilet and the lobby.

We also define a few activities to illustrate the range of possibilities that can occur in these complex, multi-agent environments. A total of 29 staff members are tracked and analyzed, with an accompanying assumption that one person (individual depicted in purple in **Figure 2**) is infectious. The behavioural patterns of the individuals can be modified through the ABM simulations.

### 3.1.1. Stationary Activity

A characteristic stationary activity involves coordinated gathering of the staff members for discussion. During such seminars and presentations, staff members are typically gathered at a common location (e.g. a large meeting room table) and remain mostly stationary during this activity. For our simulations, we assume the 29 staff to be seated around a central table (**Figure**

**2**), with one of these individuals (also designated as the infected individual in purple colour) seated in front of them.

### 3.1.2. Mobile Activities

A mobile activity can also be defined where staff members are free to move around and interact more closely. To see how an individual's choice of activity affects the risk of infection, we consider three possible different activity scenarios:

**Activity 1:** The infected agent goes to the restroom first once the break starts, optionally joins a subgroup to interact with others, and then returns to the table when the break ends.

**Activity 2:** The infected agent joins one of three subgroups to interact with others once the break starts and then returns to the table when the break ends.

**Activity 3:** The infected agent remains at the table during the break while the rest of the susceptible agents interact in the individual subgroups.

To provision for the possibility of extended restroom use in **Activity 1**, we assign one instance out of a hundred (which we term as 'extended case') that an infected agent remains in the restroom after entering it and only returns to the table at the end of the break.

## 3.2. Aerosol transport

We observed that aerosols originating from the infected individual closely follow the airflow and recirculate within the vortex formed in front of the individual. **Figure 3** visually demonstrates the concentration of emitted aerosols (i.e. the passive scalar) across the table from the infected source. Aerosol concentrations peak near the source (mouth) and decrease over distance as they disperse into the surrounding space. Accordingly, susceptible individuals nearest the source are at greater risk of airborne infection compared to those further away. More details of the airflow can be found in **Appendix C**.

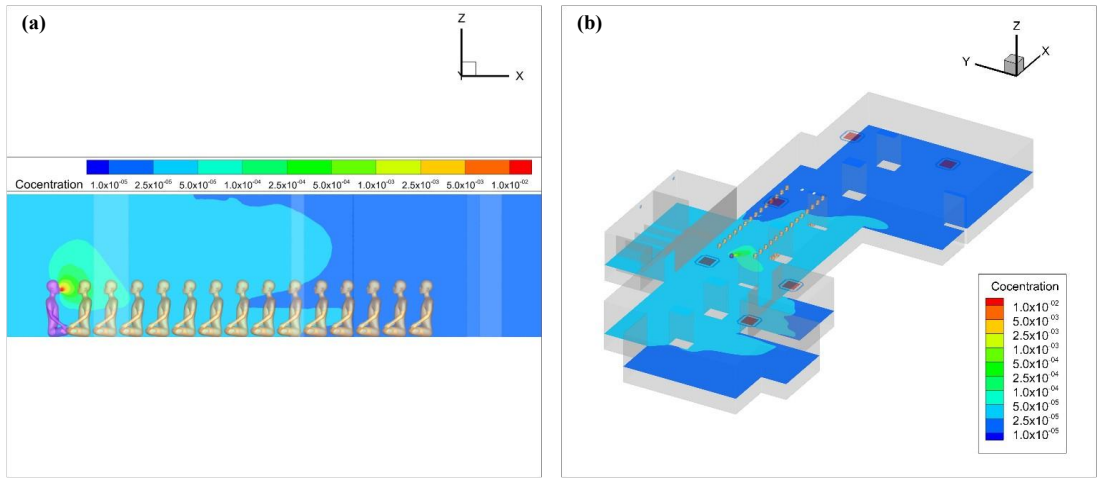

**Figure 3.** Contour plots of the passive scalar concentration at (a) a vertical cross-sectional slice across the table and (b) a horizontal slice at 0.8m above floor. The infectious individual (i.e. source, in purple) is seated at the head of the table. The other human models (in yellow) seated around the table represent susceptible individuals.

### 3.3. Agent dynamics

**Figure 4** shows a human mobility simulation following the paths taken by selected agents undergoing various simulated activities during a break, as detailed in **Section 3.1.2**. Following the blue line (Trajectory 1), the infected source may visit the washroom first, before joining a subgroup to talk with other individuals after the washroom and finally returning to the table before the break ends. Trajectories 2 and 3 are depictions of paths susceptible agents took when breaking out into the different social subgroups during the break. Three locations are marked as potential sites for the convergence of individuals to form such subgroups (pink boxes in **Figure 4**).

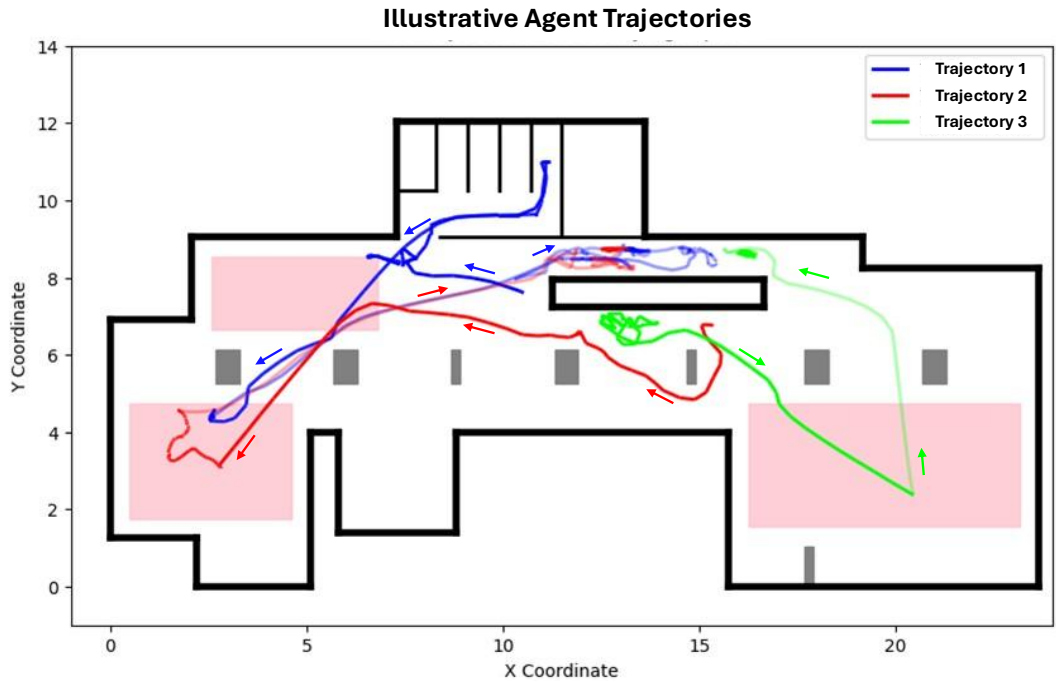

**Figure 4.** Illustration of the variety of paths that can be taken by a potentially infected individual (e.g. the blue line), and other (potentially susceptible) individuals in the human mobility simulations. 3 representative trajectories are illustrated: Trajectory 1 (in blue) shows the individual queueing and visiting the restroom prior to joining the subgroup discussions; Trajectory 2 (in red) shows the individual briefly joining the top-right subgroup group before heading to the bottom-left subgroup group; Trajectory 3 (in green) shows the individual going to a subgroup discussion at the bottom-right corner of the room. All individuals return to the table once the 15 min activity ends. Solid and fainter line segments represent earlier and later intervals in the 15 min (with arrows indicating the direction of travel for each trajectory).

### 3.4. Risk assessment

Using aerosol transport data from CFD-ML (**Section 3.2**) and agent mobility data from ABM (**Section 3.3**), we can compute the probability of infection for all human agents for variants of a simulation scenario.

#### 3.4.1. *Effect of variations in the activity of the infectious agent*

The proposed incorporation of ABM allows us to study the effects of possible activity and interactions on the overall infection probability. To reiterate briefly, Activity 1 incorporates a washroom trip by the infectious individual, Activity 2 involves a simple subgroup discussion and Activity 3 is a scenario where the infected individual does not interact with others but remains at the table alone (see **Section 3.1.2**). Each of these activities are simulated with ABM for a 15 min duration. Based on 100 random simulations for each activity scenario (1–3), the probability of infection for all susceptible agents is tallied and binned in histograms (**Figure 5**). The viral load from the infected source can vary across several orders of magnitude depending on stage of infection, and this can be customized in the simulation as desired. Here, we chose a mean emission rate $ER_q$ of 7.69 (q/min) based on one standard deviation above the mean reported viral load (see **Appendix G**).

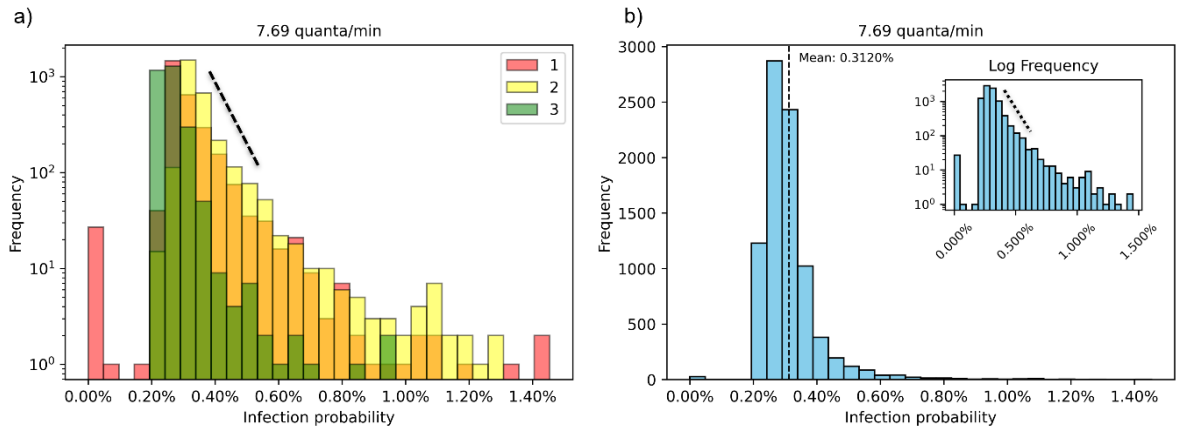

**Figure 5.** (a) Frequency counts (semi-log scale) of infection probability accumulated over the 15 min break session based on 100 simulated instances for each activity scenario, 1 to 3. (b) Total frequency count across the three activity scenarios of infection probability where the mean infection probability is approximately 0.312%. Inset on semi-log scale highlights bimodal distribution at low infection probabilities. Dashed lines are eye-guides for exponential decay of probability distribution.

Based on **Figure 5**, we highlight three interesting observations:

1. *Long-tailed distribution* – When the infected agent visits the washroom (Activity 1) or joins a subgroup discussion (Activity 2), he/she exposes other susceptible individuals to the possibility of greater risks of infection (long-tailed distribution up to 1.5%), compared to when he/she remains seated at the table (Activity 3), where the maximum probability is limited to approximately 0.95%. This is consistent with intuition, as the infectious individual's decision to visit the washroom and/or join the subgroup results in close encounters with other susceptible individuals (see **Figure 4**). The long-tailed distribution deviates from the Wells-Riley assumption of homogeneous well-mixed aerosol concentration[37] and highlights the impact of human decisions on indoor risk assessment.

2. *Bimodal distribution* – Instead of the expected unimodal distribution, the aggregated histogram (see inset of **Figure 5b**) reveals a bimodal shape, with one group of agents at moderate-to-high risk, and a second group with extremely low infection probabilities (<0.1%) from Activity 1 exclusively (**Figure 5a**). On further analysis, we found that this is a result of the infectious source agent visiting the toilet during break time, as the toilet is a negative pressure facility with continually operating exhaust fans (**Appendix H**). Because the transmission risk is effectively contained for as long as the source remains in the toilet, the observed low probability mode is dominated by agents found in the extended case where the source agent does not return to the table until the break ends.

3. *Exponential decay* – The right-skewed shape of the infection probability histogram (**Figure 5a**), particularly Activity 3, can be explained by modeling the exposure of mobile susceptible agents to a stationary infectious source using a random walk framework. This leads to exponentially distributed exposure durations, causing the resulting infection probabilities to follow an exponential distribution at low values, which leads to the approximately linear decay observed on a semi-log scale (eye-guide, **Figure 5a**). Derivation and assumptions underlying this result are detailed in **Appendix I**. In real-world terms, the random walk basis reinforces that short, sporadic contacts can cumulatively lead to meaningful exposure with time. The exponential decay trend implies that modest reductions in duration or frequency of contact can still yield significant reduction in transmission risk.

The 5th–95th percentile infection probabilities range from 0.225% to 0.468% based on aggregated data (**Figure 5b**), reflecting more than a twofold difference in risk between the least and most exposed individuals in a shared environment. Further, the 1st–99th percentile spans 0.213% to 0.695%, highlighting the presence of a long tail driven by a small subset of individuals experiencing disproportionately high exposure.

### 3.4.2. *Infection probability during training versus break*

Based on this methodology, we can flexibly deconstruct activity patterns of individuals to account for the risk factors due to heterogeneous activities. As another example of how agent movement affects infection probability, we modeled an hour-long training session composed of a 45-minute discussion at the table, followed by a 15 min subgroup discussion.

**Figure 6** shows the probability of infection for each human agent seated along the lower row (B1–B14, in blue) and upper row (T1–T14, in red). The agents are numbered in ascending sequence by their distance from the infected source (individual in purple) as depicted in **Figure 6a**. **Figure 6b** shows the accumulated risk of infection over a 45 min training period. The agents near the infected source are at a greater risk than the other agents further away, especially for agents along the lower row who are downstream of the prevailing air flow (indices: B1-3). The spatial heterogeneity due to local effects like the air-conditioning unit is also shown to cause individuals further away to have slightly higher risk, e.g. individual B8 has slightly higher risk than B7, potentially due to their positions relative to the air-conditioning unit above them.

The 45 min training period is followed by break activities 1, 2 or 3, each lasting 15 min. Depending on the activity, the actions taken directly impact the infection probability accumulated across agents (modelled by ABM), causing a further variation in total infection probability as shown in **Figure 6c–e**.

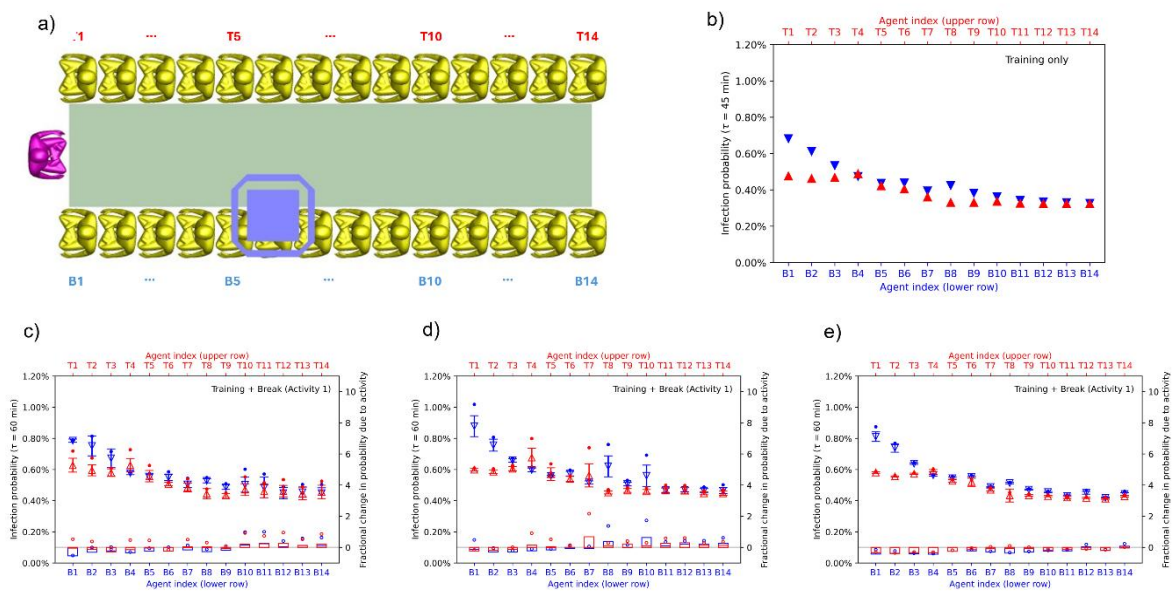

**Figure 6**. (a) Schematic of agents analyzed in (b)-(d), numbered by their placements in the upper (Top) or lower (Bottom) rows, and their relative proximity to the infected source (in purple). 1 is the nearest, while 14 is the furthest. (b-d) Probability of infection for each human agent seated along the lower row (in blue) and upper row (in red) accrued over (b) a 45 min training, and followed by a 15 min break consisting of (c) activity 1, (d) activity 2, or (e) activity 3. Error plots indicate standard deviation of the infection probability (left ordinate) and filled dots denote upper 95% of the ABM data distribution. Open bars indicate mean fractional change in infection probability due to break activity (right ordinate) and open circles denote the upper 95% of the ABM data distribution. A mean emission rate, $ER_q$, of 7.67 (q/min) is assumed for this analysis.

To compare the associated infection risks due to training and break activity, we define a fractional change in infection probability of agent j as

$$\Delta \tilde{P}_j = \frac{P_j^b(\tau_b) - P_j^t(\tau_b)}{P_j^t(\tau_b)}, \qquad (3)$$

where $P_j^b$ and $P_j^t$ the infection probabilities accumulated over break (superscript $b$) and training (superscript $t$) period $\tau_b$ of 15 min and 45 min respectively. Results are plotted in **Figure 6c–e**. On average, we see that the agents who are sitting near the infected source during training have relative infection risks reduced during break time ($\Delta \tilde{P}_j < 0$), whereas those sitting further away are at relatively greater risks during break time ($\Delta \tilde{P}_j > 0$). Note that agent mobility during break time creates opportunities for close interaction and engagement with the source. In some cases, such as agent T4 in Activity 1, the 95$^{th}$ percentile infection probability during break time *increases* the overall infection risk significantly, even though the mean infection probability shows a decrease for the same amount of time. Separately, the 95$^{th}$ percentile infection probability for agent B10 during Activity 2 is more than twice that of the training session. Animated trajectories are provided in Supplementary Videos 1 and 2, depicting how the highest percentile infection probabilities for B10 and T4 are a result of proximity between the specific susceptible agent and the infectious agent during the break period.

We conducted one-sided hypothesis tests to evaluate whether each activity altered infection risk relative to remaining seated. Activity 1 did not increase mean infection risk ($t = -1.50$, $p = 0.93$) but significantly elevated 95th percentile risk ($t = 4.14$, $p < 0.001$). Activity 2 showed weak evidence of increased mean risk ($t = 1.61$, $p = 0.06$) and a strong increase in 95th percentile risk ($t = 3.92$, $p < 0.001$). By contrast, Activity 3 significantly reduced both mean ($t = -8.50$, $p < 0.001$) and 95th percentile ($t = -4.83$, $p < 0.001$) infection risks and therefore emerges as the safest activity option.

Importantly, the modular nature of the proposed framework allows for one to account for individuals' exposure through various scenarios by flexibly composing different types of activities through the day.

## 4. Discussion

Our unified CFD-ML-ABM framework can assess indoor airborne transmission risks based on the confluence of environmental factors, human mobility and disease biology. The modular framework proposed is entirely extensible to other airborne infectious disease, activities, or venues. Using a childcare centre venue as a case study, we demonstrated how the framework could account for effects of air flow on aerosol transport and dispersion and distinguish the risk profiles of individuals who choose to undertake certain activities within an indoor working environment. Our results suggest that a risk assessment based on distancing alone neglects the local spatial heterogeneity due to the specific indoor ventilation and the frequency of interactions because of ad-hoc human decisions, aspects often understated in literature.

Notably, our model is easily configurable to accommodate different human activities. When a specific infectious agent is assumed, this analysis can also illuminate the dual impact of both a sedentary activity (discussion around a table), where specific agents nearer the infectious individual are at higher risk (greater impact from local airflow dynamics), and the mobile activity (break), where different random individuals may be exposed to higher doses due to the infectious agent's behavior. Other spatio-temporal activities, such as training sessions and group interactions, can be reconstructed based on observations or contact tracing logs and simulated to assess their cumulative impact on infection risk.

A central finding is that the infection probability distributions are not only heterogeneous but also structurally dependent on the activity. In Activity 1, we observed a clear bimodal distribution, which we traced to scenarios where the infectious source agent enters the toilet, a negative pressure zone with exhaust ventilation. In such cases, the source is effectively isolated from susceptible agents, resulting in a low-probability mode. In addition, we identify a secondary low-risk mode associated with the source agent spending time at the open toilet entrance, where partial containment reduces transmission risk. While this specific scenario was not explicitly sampled in the limited ABM runs for Activity 2, it remains a potential contributor to the low-risk mode.

In contrast, Activity 3 features a stationary source agent in the main hall who does not actively engage others. The resulting infection probability distribution is unimodal but exhibits an exponential decay, which we show arises naturally from the stochastic motion of susceptible agents relative to a fixed source. This behavior is consistent with the random walk exposure model. These findings support targeted interventions such as minimizing mingling duration during breaks, restructuring activity planning and rapidly isolating potential sources in well-mixed indoor environments.

On the other hand, the long tails observed in Activities 1 and 2 stem from sustained agent proximity and repeated interactions, which are not captured by the exponential approximation. These long tails disproportionately account for the highest transmission risks, especially in heterogeneous settings where a subset of individual are in close proximity with an unidentified source for extended periods. This suggests that even modest interventions focused on the most exposed individuals can yield disproportionately large reductions in overall transmission.

In addition to aerosol transport and human behavior, infection risk assessment depends significantly on biological inputs such as viral load and emission rates and physical activities affecting breathing volume and frequency. The emission rate $ER_q$ used can significantly affect the probability of infection as calculated. There are multiple parameters involved in the estimation of emission which can be flexibly adapted to reflect different pathogen types, or types of activities, such as singing or shouting, allowing the model to be applied across a broad range of airborne disease transmission scenarios.

Our proposed risk assessment framework can flexibly account for differences in inhalation rates from different activities, pulmonary absorption rates and protective equipment including masks, with additional scaling factors as proposed in Mittal et al.[34] The flexibility of this framework also allows for quantitative comparisons of different human activity-based interventions, e.g. the imposition of specific interaction zones or restrictions in break (interaction) duration.

There are several limitations and room for improvement in the proposed framework. The model is minimal by construction and assumes only the mean viral loadings and idealized dose responses. It does not currently account for larger droplets (typically larger than 5 μm) that settle quickly, land on a fomite such as the meeting table and cause infection by indirect contact transmission via face touching.[42] Also, it provides predictions of transmission risks based on known information and inherits all limitations from its constituent models. For example, the complexity of infectious disease biology, along with significant uncertainties in parameterization, can greatly undermine predictions of absolute risk of infection (even if relative comparisons are fairly robust). Similarly, the CFD model for airflow is idealized and minimal, steady and does not account for human agents and their movements while the ML model is reduced-order, bound by a fixed domain, and generalizes information from a three-dimensional dataset to a depth-averaged two-dimensional output (i.e. breathing height). The autonomous agents in ABM are also limited in the number of decision options, respond elastically to chance encounters and are singularly focused on executing decisions.

Notwithstanding, the proposed framework presents a useful approximation to a highly variable and complex interplay between human decisions and environmental factors. Our work adds to a growing literature on respiratory aerosol transport leading to transmission of infectious diseases,[1] indoor risk assessment,[43] and mitigation[44] in the wake of the COVID-19 pandemic. Future work can focus on streamlining the process workflow so the framework can be deployed and implemented quickly in preparation for the next global pandemic.

**Data availability**

The data that support the findings of this study are available from the corresponding author upon reasonable request.

**Code availability**

The code used to implement the CFD, ML, and ABM models is available from the corresponding author upon reasonable request.

**Declaration of interests**

All authors declare no competing interests.

# Appendix

## A. Governing equations

The motion of the air is described by governing equations for mass and momentum, which under steady state conditions, reads

$$\nabla \cdot (\rho \vec{u}) = \dot{m}, \tag{4}$$

$$\nabla \cdot (\rho \vec{u} \vec{u}) = -\nabla P + \nabla \cdot [(\mu + \mu_t)(\nabla \vec{u} + \nabla \vec{u}^T)] - \nabla \cdot \left(\frac{2}{3}\rho \kappa \vec{I}\right) + F_m, \tag{5}$$

where $\kappa$ is the turbulent kinetic energy and $\varepsilon$ is the dissipation of turbulent energy, expressed as

$$\nabla \cdot (\rho \vec{u} \kappa) = \nabla \cdot \left(\frac{\mu_t}{\sigma_k} \nabla \kappa\right) + G_k - \rho \varepsilon, \tag{6}$$

$$\nabla \cdot (\rho \vec{u} \varepsilon) = \nabla \cdot \left(\frac{\mu_t}{\sigma_\varepsilon} \nabla \varepsilon\right) + \frac{\varepsilon}{\kappa}(C_{1\varepsilon} G_k - C_{2\varepsilon} \rho \varepsilon), \tag{7}$$

where $C_{1\varepsilon}$ and $C_{2\varepsilon}$ are constants valued 1.44 and 1.92 respectively, $\sigma_\kappa$ and $\sigma_\varepsilon$ are 1.00 and 1.3 respectively, and $G_k$ represents the production of turbulence kinetic energy. Eddy viscosity $\mu_\tau$ is expressed as

$$\mu_t = \rho C_\mu \frac{\kappa^2}{\varepsilon}, \tag{8}$$

where $C_\mu$ is equal to 0.09. Conservation of energy is ensured by

$$\nabla \cdot (\rho \vec{u} E) = \nabla \cdot (\lambda \nabla T), \tag{9}$$

where $E$ is energy and it is expressed as

$$E = h - \frac{P}{\rho} + \frac{\vec{u}^2}{2}, \tag{10}$$

$h$ is the sensible heat and it is represented by

$$h = \sum_i x_i C_{p,i} T + \frac{P}{\rho}, \tag{11}$$

The transport of aerosols in the form of a passive scalar concentration $\phi$ is described by

$$\nabla \cdot (\rho \vec{u} \phi) = \nabla \cdot \vec{J} + S, \tag{12}$$

where $S$ is the source term of scalar and $\vec{J}$ is the diffusive flux of the scalar and can be expressed as

$$\vec{J} = -\left(\rho D + \frac{\mu_t}{Sc_t}\right) \nabla \phi, \tag{13}$$

where $Sc_t$ is the turbulent Schmidt number (taken as 0.7) and $D$ is diffusivity.

## B. Computational domain

We performed geometric meshing for wall boundaries of the computational domain, including air conditioning units and human models (**Supplementary Figure 1**). The mesh was generated using the cut-edge mosaic meshing scheme, which filled the bulk region with octree hexes while maintaining a high-quality, layered poly-prism mesh in the boundary layer. The minimum and maximum mesh size of the domain are set to 0.0025 m and 0.2 m respectively, with a mesh growth rate of 1.2. Different grid sizes were employed across the domain to meet specific spatial requirements. The mesh for walls is limited to a maximum grid size of 0.05 m. To ensure better resolution of the ceiling air-conditioning units, the maximum grid size is refined to 0.01 m (**Supplementary Figure 1a**). To accurately resolve aerosol transport near human agents, we further refine the mesh on human bodies to 0.005 m and mouth to 0.0025 m (**Supplementary Figure 1b**). In total, each simulation case comprises approximately 12.7 million mesh elements.

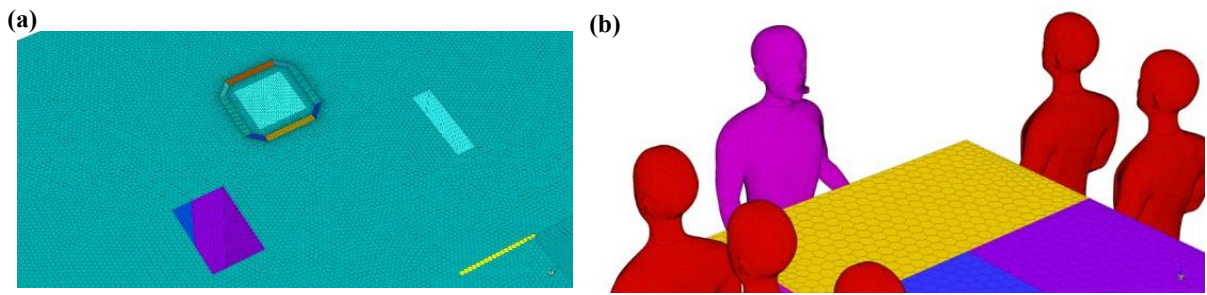

**Supplementary Figure 1.** (a) Iso-geometric view of the mesh of the ceiling air-conditioning units (denoted by blue squares in **Figure 2**). (b) Mesh for human models for the infectious individual (in purple) and susceptible individuals (in red).

The ceiling air-conditioning cassettes are air exchange units where air flows into computational domain from four air-supply slots at the sides and flows out through the square center. The total cool air supply of 0.39 m$^3$/s at 22⁰C is discharged from each cassette slot at a 30-degree downwards angle. Air, including aerosols, is extracted out of the domain with a reduced factor to account for filtration effect. The exhaust fan in the toilet is modelled as a boundary surface with a fixed flow rate of 0.1 m$^3$/s out of the computational domain. Non-slip boundary conditions apply to all wall surfaces. The surface temperature of model models is set at 36°C, while all other surfaces were assumed to be adiabatic. Aerosols are released from nose and mouth region of the infected source model, defined by a 2 cm cube as placed at 1.5m height. A normalized value of 1 is assigned as the source term in the passive scalar equation.

### C. Airflow patterns

**Supplementary Figure 2** illustrates the discharge of cool air from the AC cassettes' side openings at a 30-degree downward angle. The Coanda effect influences the trajectory of the discharged fluid jet, causing it to adhere to the ceiling walls. When it encounters side walls or obstructions, it transitions downward and recirculates back to the returns located at the center of each AC cassette. A velocity vector plot of the vertical cross-section is shown in **Supplementary Figure 3**, which reveals a vertically recirculating flow in front of the individual, primarily induced by the air-conditioning.

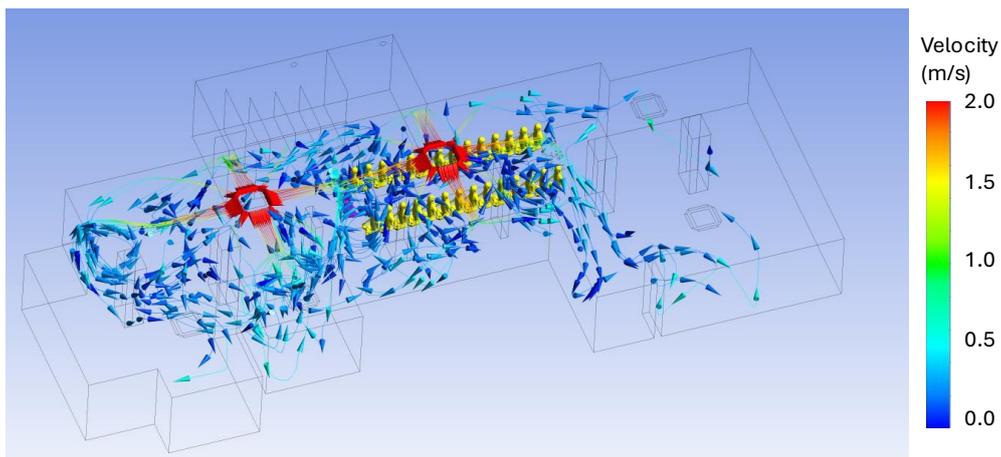

**Supplementary Figure 2.** Velocity streamlines originating from the air-conditioner units in the venue.

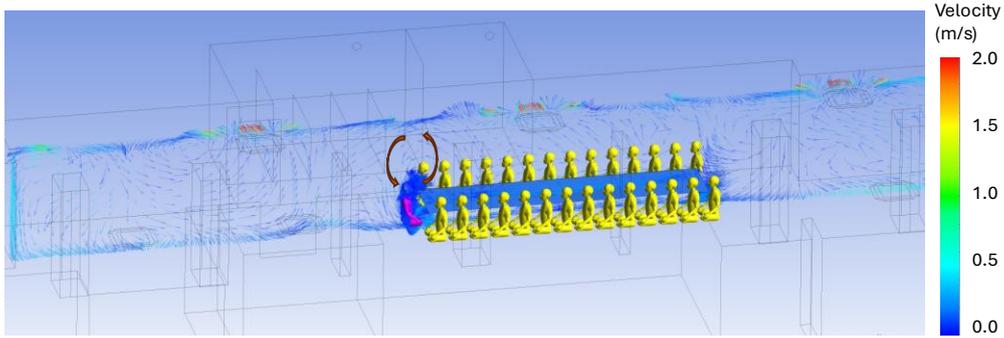

**Supplementary Figure 3.** Velocity vector plot illustrating the airflow dynamics across a vertical cross-section in the venue.

D. Dispersion and clearance

As a simplifying assumption, steady-state simulations are used for training the ML model to predict the passive scalar concentration as any receiver location within the domain, given a specific source location. Since the training data was extracted from steady-state simulations, we assume that infected individual's airborne viral emissions are spatially dispersed quickly (e.g. from a breath) and cleared before the next emission event, as shown in **Supplementary Figure 4**.

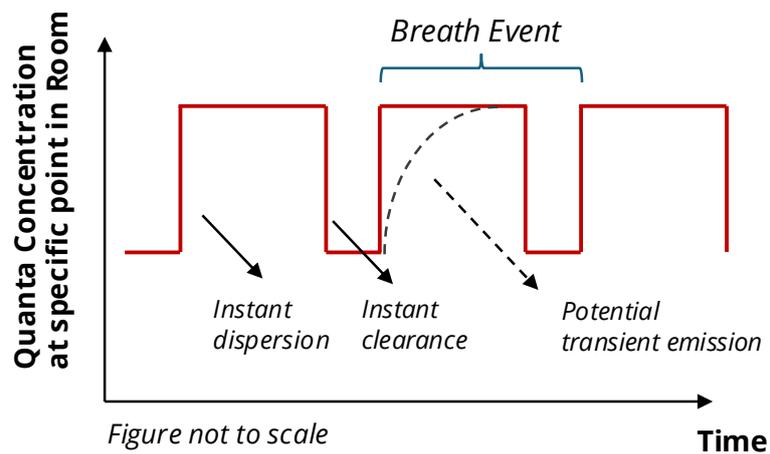

**Supplementary Figure 4.** Schematic diagram showing instant clearance and dispersion in time for ML models.

This neglects the temporal dynamics of aerosol buildup and decay, which can be crucial, especially for longer-time simulations. Nonetheless, for this particular scenario, we note that the calculated ACH, based on observations at a real facility, was approximately 19.4, which approximates to a clearance time constant of 3 min (under the well-mixed assumption). Given a simulation (activity) time of 15 minutes, this means that more than 99% of the original aerosol is cleared from the domain by the end of the 15 min. Nonetheless, a transient model may be required for much longer activities.

### E. Surrogate model training and selection

Two distinct surrogate modelling techniques were trained and tested to evaluate their relative efficacy: XGBoost and Multi-layer Perceptrons (MLPs). The first surrogate model employed, XGBoost, is a gradient boosting framework that utilizes decision trees as base learners. XGBoost was chosen for its ability to handle large datasets efficiently relative to other regression trees and has been shown to be robust and effective across multiple regression tasks in different domains. The second surrogate model, Multi-layer Perceptrons (MLPs), belongs to the class of artificial neural networks (ANNs) and is characterized by multiple layers of interconnected neurons. MLPs are common neural network architecture and among the simplest versions of the class of deep learning models that have been shown in recent years to have extremely good capability to capture complex nonlinear relationships within the data.

Both surrogate models were trained in a similar fashion, with five inputs comprising the respective (x,y) positions of the infectious agent (source) and the uninfected agent (receiver), and the horizontal distance between them. The models' output is a single value corresponding to the normalized passive scalar concentration at a specific location (putatively corresponding in turn to the expected high-fidelity simulation's concentration). While the numerical simulations provided complete 3D passive scalar concentrations, values at a single cut-plane were specifically extracted for the purposes of the surrogate model to correspond to putative inhalation and emission at standing height. In total, 100 locations within the indoor space.at 1.5 m height (corresponding to an average adult's breathing zone height) were selected using a Latin Hypercube Sampling (LHS) scheme for data generation.

This can be mathematically represented in the following form:

$$c(x_{receiver}, y_{receiver}) = f_{surrogate}(x_{source}, y_{source}, x_{receiver}, y_{receiver}, dist_{2D}). \quad (14)$$

The trained models were separately assessed for their predictive accuracy and generalization capability via a Leave-One-Out-Cross-Validation process. We use mean squared error (MSE) and mean absolute percentage error (MAPE) metrics to evaluate and compare the error between the two surrogate models. Briefly, the MSE metric quantifies the average squared difference between the predicted and actual concentration values in the dataset whereas the MAPE metric quantifies the average absolute percentage difference between the predicted and actual values of a variable. As the passive scalar concentration spans multiple orders of magnitude across the entire spatial domain (depending on distance from the source), we further employ a log-transform on the output to bring the regression outputs to a similar scale. Hence, the MSE and MAPE metrics are also reported on the regression outputs on a log-scale.

**Table 1.** Performance of surrogate models in terms of mean squared error (MSE) and mean absolute percentage error (MAPE).

| Model | MSE | MAPE |
| --- | --- | --- |
| XGBoost | $1.8 \times 10^{-4}$ | 0.82% |
| MLP | $1.6 \times 10^{-4}$ | 0.91% |

Overall, the results indicate that the 2 different surrogate models perform similarly well in the context of this work. **Supplementary Figure 5** shows that the spatial distribution of the predicted aerosol concentrations within the childcare center are qualitatively well-captured by both surrogate models. For simplicity, we chose the XGBoost model for integration with the ABM models.

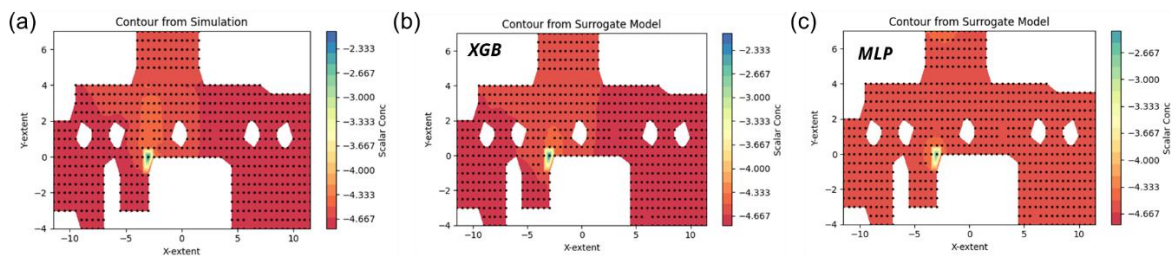

**Supplementary Figure 5.** Contour plots of aerosol concentrations in passive scalar (colorbar in log scale) in the childcare center obtained from (a) the high-fidelity numerical simulations, and surrogate models (b) XGBoost and (c) MLP.

There is potential scope for extending this framework to the modelling of transient dispersion dynamics as well.

F.  **Modelling agent mobility**

Under the ABM framework (**Section 2.4**), the stepwise movements of a human agent are described based on the social force model.[45] The position $\vec{x}_i(t)$ and velocity $\vec{v}_i(t)$ of an individual agent $i$ at time $t$ are described by equations of motion as

$$\frac{d\vec{x}_i(t)}{dt} = \vec{v}_i(t), \tag{15}$$

$$\frac{d\vec{v}_i(t)}{dt} = \vec{f}_{i,d} + \sum_{j \neq i} \vec{f}_{ij} + \sum_B \vec{f}_{iB}, \tag{16}$$

where the driving force term $\vec{f}_{i,d}$ reflects the tendency of individual $i$ to move toward their destination. The interpersonal repulsive force term $\vec{f}_{ij}$ describes the tendency to maintain a certain distance from other individual j, and $\vec{f}_{iB}$ is for the repulsive interaction with boundary $B$, such as walls and obstacles.

The driving force $\vec{f}_{i,d}$ is

$$\vec{f}_{i,d} = \frac{v_d \vec{e}_i - \vec{v}_{i(t)}}{\tau}, \tag{17}$$

where $v_d$ is the desired walking speed and $\vec{e}_i$ is desired walking direction vector, a unit vector pointing the destination of individual $i$. The relaxation time $\tau$ reflects how fast individual $i$ changes its velocity to the desired velocity.

The interpersonal repulsive force term $\vec{f}_{ij}$ is

$$\vec{f}_{ij} = C_A \exp\left(-\frac{d_{ij}(t) - r_i - r_j}{C_B}\right)\vec{e}_{ij} + \vec{g}_{ij}, \tag{18}$$

where $C_A$ and $C_B$ are the scaling parameters for the strength and distance of interpersonal repulsive force, respectively. The distance between individuals $i$ and $j$ at time $t$ is indicated as $d_{ij}(t)$, and $\vec{e}_{ij}$ is a unit vector pointing from $i$ to $j$. The radii of individual models $i$ and $j$ are denoted by $r_i$ and $r_j$.

The physical interaction force term $\vec{g}_{ij}$ indicates body compression and friction between individuals $i$ and $j$

$$\vec{g}_{ij} = h\left(r_{ij} - d_{ij}(t)\right)\{k_n \vec{e}_{ij} + k_t[(\vec{v}_j - \vec{v}_i) \cdot \vec{t}_{ij}]\vec{t}_{ij}\}, \tag{19}$$

where the function $h(x)$ returns $x$ if $x > 0$, otherwise 0. A unit vector $\vec{t}_{ij}$ is perpendicular to $\vec{e}_{ij}$, indicating the direction of friction force. the normal and tangential elastic constants are indicated by $k_n$ and $k_t$, respectively. Similarly, the boundary repulsive force term $\vec{f}_{iB}$ is given as

$$\vec{f}_{iB} = C_A \exp\left(-\frac{d_{iB}(t) - r_i}{C_B}\right) \vec{e}_{iB} + \vec{g}_{iB}, \tag{20}$$

where $d_{iB}(t)$ is the distance between individual $i$ and the boundary $B$, and $\vec{g}_{iB}$ for the physical interaction force between them.

Our human mobility model is implemented in MomenTUMv2 software developed by Kielar et al.[45,46] The parameter values of the operational layer in **Table 2** are selected based on previous works.[45,46]

Table 2. Operational layer parameter values [45,46]

| Parameter | Symbol | Value |
| --- | --- | --- |
| desired speed | $v_d$ | 1.34 m·s$^{-1}$ |
| relaxation time | $\tau$ | 0.5 s |
| strength scale parameter | $C_A$ | 20 m·s$^{-2}$ |
| distance scale parameter | $C_B$ | 0.04 m |
| individual radius | $r_i$ | 0.2 m |
| normal elastic constant | $k_n$ | 14000 m·s$^{-2}$ |
| tangential elastic constant | $k_t$ | 2000 s$^{-1}$ |

### G. Estimating quanta emission rate

Here we estimate the mean quanta emission rate from a source under light activity. The quanta emission rate $ER_q$ (q/min) is defined as,

$$ER_q = c_i \cdot c_v \cdot V_d \cdot IR, \tag{21}$$

where $c_v$ is the viral load (RNA/ml) in the range $10^3 - 10^{11}$,[47] broadly categorized as $10^7 - 10^9$ for mild to moderate cases and $10^9 - 10^{11}$ for moderate to severe cases.[48] Here we use a mean viral load of $10^{8.57}$ based on an upper standard deviation bound of reported viral loading frequencies for SARS-Cov-2,[49] as shown in **Supplementary Figure 6**. The conversion factor $c_i$ is the dose response of one infectious quantum per viral copy (q/RNA). Here we estimate $c_i \approx 10^{-4}$ (q/RNA) based on dose response estimate of 1/12 (q/pfu),[50] and assay conversion of 1/822 (pfu/RNA).[51] Depending on the nature of expiratory activity, droplets are expelled from the host at droplet volume concentration $V_d$ (ml/L), ranging from $2 \times 10^{-6}$ (oral breathing) to $6 \times 10^{-5}$ (vocalizing). Here we estimate $V_d \approx 9 \times 10^{-6}$ (ml/L), assuming talking activity.[47] $IR$ is the ventilation rate (L/min) estimated as 23 (L/min) assuming walking activity.[47]

Together, the mean quanta emission rate from a walking, talking source is estimated as $ER_q \approx 7.69$ (q/min).

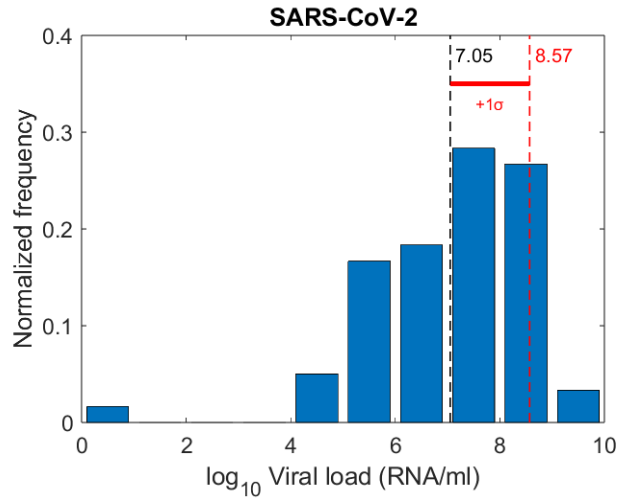

**Supplementary Figure 6.** Normalized viral load frequency distribution based on reported values.[49]

### H. Containment effect of negative pressure room

The floor plan of the childcare center (**Figure 2**) includes a toilet equipped with air vents that evacuates air from the facility, effectively acting as a negative pressure room.[52] Here, we set up a rectangular Cartesian grid of the domain with inaccessible spaces excluded and set up 100 instances where in each instance, an infected source agent is positioned at a random accessible point on the grid. If an infected source agent were far from the toilet entrance, the emitted aerosols disperse in space and time until a homogeneous well-mixed concentration[37] is achieved at the far field (**Supplementary Figure 7a**). If the source agent were standing at an open toilet entrance, a portion of the emitted aerosols could be drawn into the low-pressure toilet, leading to significantly reduced concentrations in the main hall (**Supplementary Figure 7b**). Accordingly, if the source agent were inside the toilet, the aerosols are effectively contained and the airborne transmission risk presented other agents within the main hall is extremely low (**Supplementary Figure 7c**).

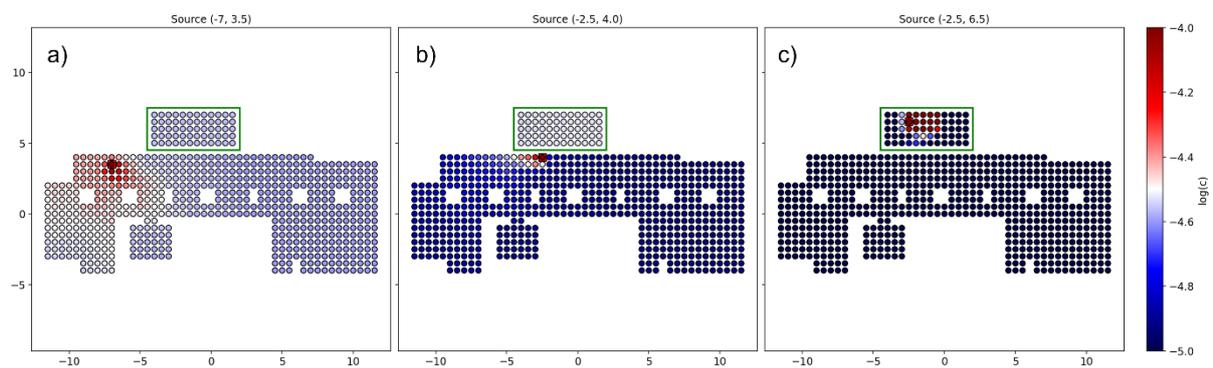

**Supplementary Figure 7. Scalar concentrations (on log scale) on a regular grid (dimensions in meters) based on floor plan of childcare center (Figure *2*), where a static source is located (a) at far away from, (b) at an open entrance of, or (c) inside the toilet (green box). Red and blue indicate high and low concentrations respectively (colormap). The toilet's negative pressure setup draws air containing aerosols away from the room, acting as a containment mechanism.**

The containment effect due to the source agent being either near to the toilet entrance or within the toilet, effectively severs the long-range airborne transmission pathway from the other agents. **Supplementary Figure 8a** shows the scalar concentration at grid points at given distance of the source, excluding instances where the source is inside the toilet. We see that the scalar concentrations are close to homogeneous over long distances which is consistent with the well-mixed assumption. In the absence of source containment, the lower standard deviation bound of the scalar concentrations remains above a minimum threshold at all distances.

To demonstrate the effect of source containment, we plot the frequency of the infection probability over 15 min exposure to the same quanta emission rate (**Section 3.4.1**) where instances of source agent in toilet are excluded as shown in **Supplementary Figure 8b.** When instances of source agent inside the toilet and at an open toilet entrance are both excluded, the three leftmost bars (in white) of the histogram vanish completely, which confirms that the low probability mode is the result of source containment. The white bars represent instances of partial containment where the source agent is at the open toilet entrance.

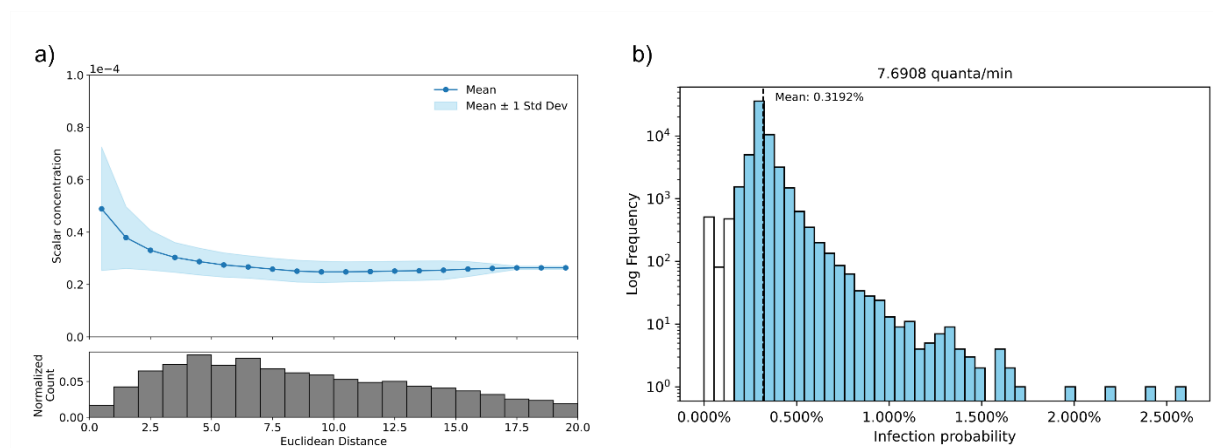

**Supplementary Figure 8.** (a) Mean scalar concentrations at grid points of given distance (in metres) from infectious source agents randomly positioned within facility, excluding the toilet. Shaded region indicates plus or minus one standard deviation from the mean values. Bottom: Normalized histogram count. (b) Frequency count (semi-log scale) of infection probability accumulated over 15 min exposure to emission rate of 7.69 quanta/min. White bars represent low-risk events, which disappear when instances of source at the toilet entrance are excluded.

I. Probabilistic decomposition of infection risk via source location and exposure time

We propose a mathematical framework to explain the observed bimodal and skewed distributions of infection risk using (1) Bayesian conditioning on source agent location and (2) stochastic modeling of exposure time.

Let $p := \mathbb{P}(\mathcal{E} = 1)$ be the probability of infection for a susceptible agent in a given scenario, here a $T = 15\ min$ break period, $\mathcal{E}$ is binary variable indicating success or failure of infection, $S \in \{IT, TE, OT\}$ is a set category indicating the location of source agent where $IT$ denotes inside toilet, $TE$ denotes toilet entrance and $OT$ denotes outside toilet, $A_k$ is a set of mobile activities (**Section 3.1.2**) simulated by ABM, where $k \in \{1,2,3\}$.

We model the distribution of probability of infection as Bayesian mixture over source locations.

$$f(p|A_k) = \sum_{s \in \{IT,TE,OT\}} \mathbb{P}(S = s|A_k) \cdot f(p|S = s, A_k), \tag{22}$$

This structure explains the observed differences in probability distributions (**Figure 5a**) across activity types:

**Activity 1:** Includes high values of $\mathbb{P}(S = IT|A_1)$ where $f(p|S = IT) \approx \delta(p)$ and $f(p|S = OT, A_1)$ is broad and long-tailed due to ABM-mediated close contacts. The resultant mixture $f(p|A_1)$ is therefore bimodal with a sharp spike at near zero $p$ and right-skewed distribution.

**Activity 2:** Includes non-zero $\mathbb{P}(S = TE|A_2)$ and long tailed $f(p|S = OT, A_2)$ due to ABM-mediated close contact. The resultant mixture $f(p|A_2)$ is unimodal with right-skewed distribution.

**Activity 3:** Includes $\mathbb{P}(S = OT|A_3) = 1$ since the source agent is immobile, which means there is no containment $\{S = IT, TE\})$ and no long tailed $f(p|S = OT, A_3)$ features since there are no active ABM-mediated close contacts. The resultant $f(p|A_3)$ is unimodal with relatively low variance.

It is important to note that the confinement effect is time-weighted over all instances. For Activity 1, the probability distribution is an ensemble average over $n$ instance of $A_1$ as

$$f(p|A_1) = \bar{\alpha} f_C(p) + (1 - \bar{\alpha}) f_N(p), \tag{23}$$

where $\bar{\alpha} := \bar{\tau}/T$ is the fraction of mean time spent in toilet $\bar{\tau}$ across $n$ instances, $f_C(p)$ is the confinement mode and $f_N(p)$ is the normal mode. The midpoint of this distribution lies in equal time-weights of confinement and normal modes, i.e. $\bar{\alpha} = 1/2$, where the source agent spends half the break time in the toilet. Based on ABM simulations (excluding extended cases), the mean time spent in toilet is $\bar{\tau} = 1.23 \ min$ with standard deviation $\sigma_\tau = 0.126 \ min$, so normal mode dominates $f(p|A_1) \approx (1 - \bar{\alpha}) f_N(p)$. For extended cases, $\bar{\alpha} \gg 1/2$, the confinement mode $f^*(p|A_1) \approx \bar{\alpha} f_C(p)$ manifests in the ensemble as a low probability peak in **Figure 5a**.

Next, we examine the shape of $f(p|S = OT, A_3)$ where the infected source agent is stationary and susceptible agents move freely. Here we show that this case can be approximated using a stochastic exposure model, which explains the exponential decay observed as a linear trend in the histograms (**Figure 5**).

Consider a susceptible agent at $X(t)$ at time $t \in (0, T]$ in bounded domain $\Omega \subset \mathbb{R}^2$. Assume a stationary source agent fixed at position $x_0$ and surrounded by an infectious circular zone

$C_r(x_0)$ of radius $r$. The cumulative exposure time of the random walker within the infectious circle is

$$\tau_r(x_0) := \int_0^T \mathbf{1}_{X(t) \in C_r(x_0)} dt. \tag{24}$$

**Postulate 1: (Exponential exposure time near fixed source)** *If the agent's motion is a time-homogeneous Markov process within a bounded domain and the infectious radius $r$ is small, then exposure time $\tau_r$ in $C_r(x_0)$ during a finite interval $(0, T]$ follows approximately an exponential distribution,*

$$f_{\tau_r}(t) \approx \lambda e^{-\lambda t}, \tag{25}$$

*for an escape rate $\lambda > 0$, assuming re-entries to $C_r(x_0)$ are negligibly rare.*[53]

**Lemma 1: (Infection probability induced by exponential exposure time)** *If exposure time is exponentially distributed and infection probability depends exponentially on exposure time (e.g. Wells-Riley model), then the infection probability is also exponentially distributed to leading order.*

Using Wells-Riley model (**Equation (1)**), the exposure time is

$$t(p) = -\frac{1}{q} \log(1 - p), \tag{26}$$

Where $p$ is infection probability and $q$ is quanta dose rate. Taking derivative,

$$\frac{dt}{dp} = \frac{1}{q(1-p)}. \tag{27}$$

Now the infection probability distribution can be obtained from the time exponential distribution (see **Postulate 1**) via change of variables as follows,

$$f_p(p) = f_T(t(p)) \cdot \frac{dt}{dp} = \frac{\lambda \exp\left(\frac{\lambda}{q} \log(1-p)\right)}{q(1-p)} = \frac{\lambda}{q} (1-p)^{\frac{\lambda}{q}-1}. \tag{28}$$

For small values of $p$, $e^{-\alpha p} \approx (1-p)^\alpha$, we obtain

$$f_p(p) \approx \frac{\lambda}{q} \exp\left(-\frac{\lambda}{q} p\right). \tag{29}$$

This exponential form explains the linear decay observed on a semi-log plot of the infection probability histogram (**Figure 5a**). This approximation is valid when exposure durations are brief and interactions are sparse, but does not capture long-tailed behavior from sustained contact (Activities 1 and 2), nor nonlinear decision dynamics modeled in the ABM (**Section 2.4**).

**Supplementary Video 1.** Animation of a specific instance of path trajectories that were taken by a potentially infected individual (i.e. the blue line), and two other (potentially susceptible) individuals in a simulation of Activity 1 (for full 15 min). Trajectory 1 (in blue) shows the individual queueing and visiting the restroom prior to joining the subgroup discussions; Trajectory 2 (in red) shows individual T4 joining the top subgroup group and being in fairly close proximity to the infectious individual (hence the relatively higher infection probability); Trajectory 3 (in green) shows individual T3 going to a subgroup discussion at the bottom-right corner of the room from the start of the break. The relative distance from the infectious individual throughout the break also results in T3 having relatively lower infection probability for this simulation.

**Supplementary Video 2.** Animation of a specific instance of path trajectories that were taken by a potentially infected individual (i.e. the blue line), and two other (potentially susceptible) individuals in a simulation of Activity 2 (for full 15 min). Trajectory 1 (in blue) shows the individual queueing and visiting the restroom prior to joining the subgroup discussions; Trajectory 2 (in red) shows individual B10 to be in close proximity to the infectious individual for extended periods during the break (hence the relatively higher infection probability); Trajectory 3 (in green) illustrates how the relative distance from the infectious individual throughout the break results in individual B9 having relatively lower infection probability for this simulation.